%% file: main.tex
\pdfoutput=1

\documentclass[conference]{IEEEtran}
\ifCLASSINFOpdf
\else
\fi
\input{packages.tex}
\allowdisplaybreaks
\input{commands.tex}
\usepackage[top=0.7in, left=0.7in,right=0.7in, bottom=0.95in]{geometry}

\allowdisplaybreaks[4]

\begin{document}
%
\title{Boosting 5G mm-Wave IAB Reliability with Reconfigurable Intelligent Surfaces}

\author{\IEEEauthorblockN{Paolo Fiore$^*$, Eugenio Moro$^*$, Ilario Filippini$^*$, Antonio Capone$^*$, and Danilo De Donno$^\mathsection$}
\vspace{-0.2cm}
\IEEEauthorblockA{
\hfill\\
$^*$ANTLab - Advanced Network Technologies Laboratory, Politecnico di Milano, Milan, Italy \\
$^\mathsection$Milan Research Center, Huawei Technologies Italia S.r.l, Milan, Italy\\
Email: $\{$paolo.fiore, eugenio.moro, antonio.capone, ilario.filippini$\}$@polimi.it, danilo.dedonno@huawei.com
}
\vspace{-10mm}
}


\maketitle

\begin{abstract}
\input{content/0_abstract}
\end{abstract}
\IEEEpeerreviewmaketitle

\vspace{-1.5mm}
\section{Introduction}
\input{content/1_intro}

\vspace{-1.85mm}
\section{Related Works}
\label{sec:ris_related}
\input{content/2_related}


\vspace{-0.3em}
\section{RIS-Enhanced Coverage Model}
\label{sec:ris_model}
\input{content/4_ris_model}

\section{Results}
\label{sec:results}
\input{content/5_results}

\vspace{-2mm}
\section{Conclusion}
\label{sec:conclusion}
\input{content/6_conclusion}
\vspace{-3mm}
\section*{Acknowledgment}
The research in this paper has been carried out in the framework of
Huawei-Politecnico di Milano Joint Research Lab. The Authors
acknowledge Huawei Milan research center for the
collaboration.

\vspace{-2mm}
\bibliographystyle{IEEEtran}
\bibliography{biblio.bib}

\clearpage
\newpage
\section{Additional material}
\input{content/7_additional_material}

\end{document}

%% file: packages.tex
\usepackage{xcolor}
\usepackage{cite}
\usepackage{amsmath}
\usepackage{amsfonts}
\usepackage[short]{optidef}
\usepackage{subfig}
\usepackage[pdftex]{graphicx}
\usepackage{autobreak}

%% file: commands.tex
\newcommand{\T}{\mathcal{T}}
\newcommand{\C}{\mathcal{C}}

%% file: content/0_abstract.tex
The introduction of the mm-Wave spectrum into 5G NR promises to bring about unprecedented data throughput to future mobile wireless networks but comes with several challenges.
Network densification has been proposed as a viable solution to increase RAN resilience, and the newly introduced Integrated-Access-and-Backhaul (IAB) is considered a key enabling technology with compelling cost-reducing opportunities for such dense deployments. Reconfigurable Intelligent Surfaces (RIS) have recently gained extreme popularity as they can create Smart Radio Environments by EM wave manipulation and behave as inexpensive passive relays. However, it is not yet clear what role this technology can play in a large RAN deployment. With the scope of filling this gap, we study the blockage resilience of realistic mm-Wave RAN deployments that use IAB and RIS. The RAN layouts have been optimised by means of a novel mm-Wave planning tool based on MILP formulation. Numerical results show how adding RISs to IAB deployments can provide high blockage resistance levels while significantly reducing the overall network planning cost. 

%% file: content/1_intro.tex
Pushed by an increasingly connected society, mobile data demands are undergoing an unprecedented growth phase that cellular networks based on sub-6GHz communications are struggling to keep up with~\cite{rappaport2019}.
To address this issue, the relatively untapped millimetre wave (mm-Wave) spectrum portion has been recently standardized by 3GPP in Release 16. Additionally, given the larger available bandwidth and the high spectral efficiency of NR, the goal of multi-gigabit mobile throughput seems to be finally reached.
However, mm-wave communications present a fundamental limitation when applied to the typical deployment scenarios of Radio Access Networks (RAN). Indeed, given the adverse propagation due to the increased carrier frequency, opaque objects such as people, vehicles and foliage interrupting the communication line of sight (LOS) between base stations and UEs might cause outages. As these objects are densely distributed in typical urban scenarios, the problem of ensuring RAN resilience against LOS obstructions becomes critical. When a radio link is blocked, not much can be done but to connect the UE to a backup base station, which must be in sight with the terminal. Consequently, the problem of guaranteeing RAN resilience has to be addressed at the system level by planning an adequate base station deployment that ensures multiple connectivity to the UEs. In particular, base station densification can naturally provide a higher reliability to high-frequency (HF) RAN while also increasing throughput~\cite{kulkarni2014}. However, denser deployments come with additional challenges, especially in terms of installation and operation costs.\newline
Among the technologies that can potentially address the issues above, Reconfigurable Intelligent Surfaces (RIS) are gaining vast popularity within the academic and industrial communities. An RIS is described as a quasi-passive planar surface that, if properly controlled, can apply several electromagnetic manipulations to impinging radio waves~\cite{Renzo2020}. In the context of cost-effective HF RAN reliability, RISs show the capability to redirect and focus impinging radio waves towards arbitrary directions, effectively behaving as passive relays. Moreover, prototypal works~\cite{tan2018} have shown that realizing RIS requires inexpensive materials, hinting that such technology can be exploited to create backup radio paths between base stations and UEs at a relatively low cost.
While the aforementioned passive-relaying behaviour of RIS has been extensively analyzed at the link level~\cite{Wang2019}, less has been done to study the integration of RIS at the system level, with a focus on boosting HF large-scale RAN reliability.\newline
This work aims at filling this research gap by analyzing the resiliency-boosting potential of RISs when employed in HF RAN. Moreover, we have included Integrated Access and Backhaul (IAB) in our analysis. IAB is a technology that minimizes dense mm-wave deployment costs by using in-band backhauling to relay access traffic. IAB is considered as a critical enabler for a cost-effective HF RAN~\cite{polese2020} and, as such, we argue that a comprehensive analysis cannot refrain from including this technology.
As previously mentioned, RAN reliability is sensitive to the final position of the installed network equipment. For this reason, we have developed a network planning model where the IAB nodes and RISs deployment over a geographic area is optimized to maximize the final RAN blockage resiliency. 

To test the reliability of the planned network during regular operation, opaque obstacles are randomly dropped into the geographic area. Furthermore, a self blockage area is added to each UE, which represents the blocking effect of the human body holding the terminal~\cite{tianyang2014}. For different obstacle densities, final results are gathered mainly by analyzing the number of UEs still retaining connectivity after the obstacle drop. This resilience indicator is evaluated for RIS-enabled IAB networks and for a baseline RAN approach where only IAB is employed. Results show that while RISs cannot substitute IAB nodes completely, the two technologies work well together to maintain high levels of network resilience, with deployment costs lower than those of IAB-only planning.
In particular, we show how RISs are most effective in boosting RAN reliability under strict budgets, confirming the cost-cutting potential that this technology is expected to offer.

The remainder of this paper is organized as follows. Section~\ref{sec:ris_related} presents some works related to IAB, RIS and mm-wave RAN. Section~\ref{sec:ris_model} describes the RAN system and the planning model. In section~\ref{sec:results}, we present numerical results coming from a reliability analysis of RAN layouts produced employing the planning optimization model. Section~\ref{sec:conclusion} concludes the paper with some final remarks.

%% file: content/2_related.tex
Despite the abundant literature on wireless multi-hop network planning started more than a decade ago \cite{AMALDI20082159, capone2010deploying}, 
a complete planning optimization model for mm-Wave IAB RANs considering capacity and reliability issues in the presence of obstacle blockages is still missing.

RISs are a very recent technological outcome \cite{gacanin2020wireless,wu2019towards}. Nevertheless, they are the main topic of an increasing number of recently published papers. They have emerged as a more energy-efficient and less costly alternative to traditional decode-and-forward approaches \cite{huang2019reconfigurable, di2020reconfigurable, bjornson2019intelligent}. Therefore, the main research direction to date has aimed to evaluate their link-level performances to establish the best RIS configuration.

Only a few works consider the impact of RISs at a network level, particularly addressing the fundamental problem of obstacle blockages. Authors in \cite{kishk2020exploiting} propose an analysis based on stochastic geometry to study the relationship between the density of RISs, network devices, obstacles and network availability. In~\cite{jarrah2021}, authors derive the symbol error probability in a mm-wave wireless mesh network that makes use of RIS.
Some of the authors of this article propose in \cite{Moro2021} a first optimization model to plan mm-Wave 5G RANs considering the RIS placement. This article extends the work by considering a different network architecture based on the IAB paradigm, the effect of obstacle blockages, and a new planning approach that leverages multi-connectivity and guarantees spatial diversity among access links.

To the best of our knowledge, this is the first work proposing a complete optimization model \textit{jointly} \textbf{1)} addressing the network planning of mm-Wave IAB radio access networks, \textbf{2)} including the RIS placement and \textbf{3)} improving the network reliability in response to sudden obstacles. Thanks to this model we can shed light on how RISs can be effectively integrated at RAN level, thus providing several hints on their final design. 

%% file: content/4_ris_model.tex
\begin{figure}
    \centering
    \includegraphics[width=0.95\columnwidth]{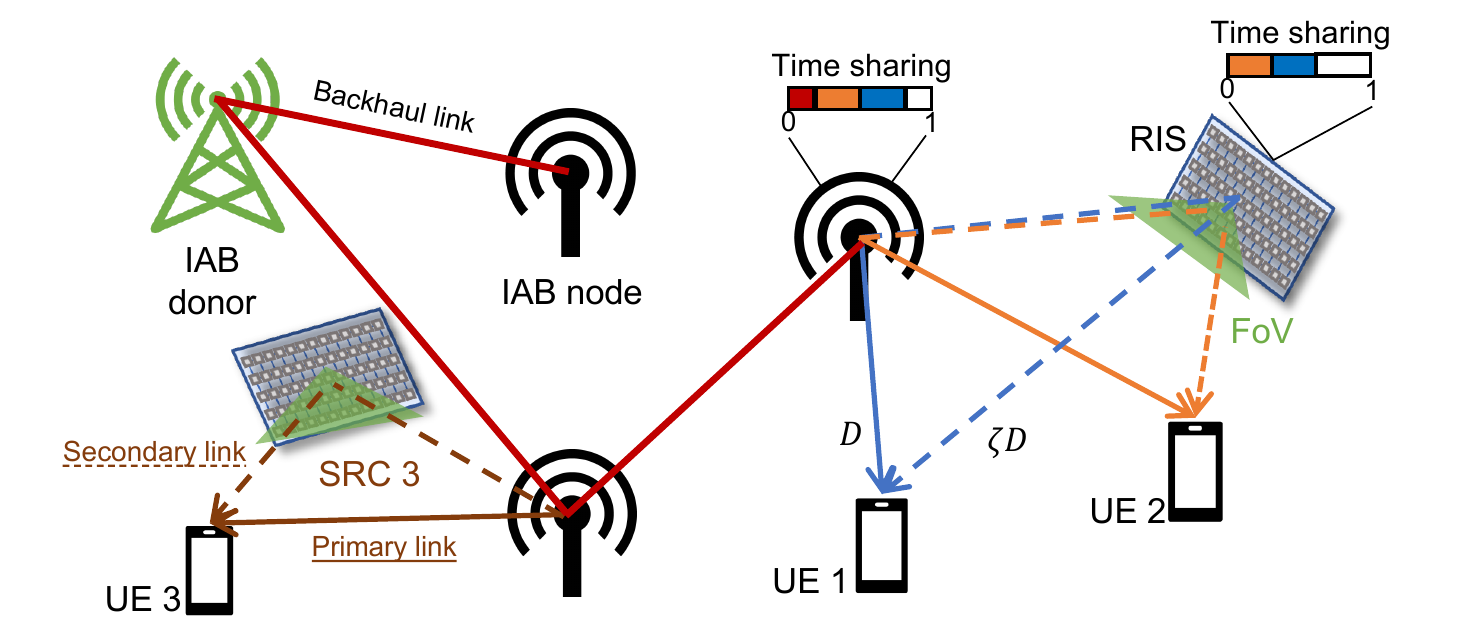}
    \vspace{-3mm}
    \caption{\small RAN scenario.}
    \label{fig:scenario}
    \vspace{-6mm}
\end{figure}

In this section, we detail the RAN system, shown in Fig.~\ref{fig:scenario}, and planning model that we employ to generate optimal RAN layouts, which will then be analyzed in terms of blockage resiliency in Section~\ref{sec:results}.
We consider RISs as \textit{passive beamformers}, capable of focusing an impinging radio wave towards arbitrary directions. This capability can be exploited at a network level to create a reflected radio path between a transmitter and a receiver, since RISs effectively behave as passive relays. The tuple comprising a transmitter, a receiver and a RIS assisting the communications is known as Smart Radio Connection (SRC)~\cite{Moro2021} and extends the well-known concept of smart radio environment~\cite{di2019smart}.
An SRC can naturally provide dual-connectivity, namely connecting the UE through multiple radio links, which is effective in reducing outages due to obstacle obstructions in mm-Wave access networks~\cite{devoti2020}.
Indeed, a RIS can be installed and configured such that it creates an alternative reflected radio path, which can be exploited to ensure continuity of service in case of direct link obstructions\footnote{Note that, since IAB donors/nodes and RISs are installed higher than UEs, backhaul and BS-RIS links are intrinsically more stable. Therefore, we do not consider BS-RIS blockages, apart from those given by fixed obstacles populating the geographic area.}, thus enhancing the mm-Wave network reliability against blockages.
However, simply deploying enough reconfigurable surfaces and base stations in a geographic area is not sufficient to guarantee good blockage resilience since other geometrical factors need to be considered, namely angular separation and link length~\cite{devoti2020}.
Indeed, a single but sizeable nomadic obstacle could interrupt both primary and backup links if these were to present low angular separation, leading to an outage and decreasing the solution's efficacy. Moreover, a significant angular separation between primary and reflected links can mitigate the well-known self-blockage effect caused by the body of the device holder interrupting the line of sight~\cite{tianyang2014}.
Link length represents another factor influencing the network resilience that can be controlled during the network planning phase. Indeed, longer radio connections naturally experience higher probabilities of being blocked by one or more obstacles~\cite{Akdeniz2014}. The proposed coverage planning formulation considers both link lengths and angular separation to maximize the network resilience against nomadic obstacles and self-blockage events. 

We now detail the assumptions and parameters that characterize the proposed planning formulation. In the following text, parameter indexes are expressed enclosed in brackets to differentiate them from optimization variables.
Similarly to what has been done in other coverage planning works~\cite{AMALDI20082159,devoti2020},
we consider a set $\C$ of Candidate Sites (CSs) over a geographic area covered by a mm-Wave access service. Each candidate site represents the position where an IAB donor (i.e., a base station with a wired connection to the core network), an IAB node (i.e., a base station with a wireless backhaul realy connection) or an RIS can be installed. A set of Test Points (TPs) $\T$ represents the UE distribution over the same area.
Physical parameters and propagation characteristics, such as transmit power, receiver sensitivity and attenuation losses, are captured by a binary activation link parameter $\Delta_{(t,c,r)}^\text{SRC}$, which can be pre-computed for all $(t,c,r)$ tuples. This is equal to 1 only when an SRC can be established between TP $t \in \T$, a donor or an IAB node installed in CS $c \in \C$ and an RIS in CS $r \in \C$. Moreover, fixed obstacles across the geographic area interrupting any line of sight can be modelled by setting the corresponding link activation parameter to 0.
Similarly, binary parameter $\Delta_{(c,d)}^\text{BH}$ indicates whether two IAB nodes in CSs $c,d\in \C$ can establish a radio backhaul link.
TP and CS are fixed, and thus achievable downlink rates in backhaul and access can be computed according to radio device characteristics and channel models. The extreme directivity of mm-Wave downlink transmissions limits the interference effects, and we assume these achievable rates to be unaffected by simultaneous transmissions\cite{devoti2020}. For each SRC $\left<t,c,r\right>$, parameter $C_{(t,c,r)}^{\text{DIR}}$ denotes the achievable TP $t$ downlink rate obtained through the direct link with the IAB node $c$, namely the direct link capacity. On the other hand, $C_{(t,c,r)}^{\text{REF}}$ indicates the achievable rate when the RIS-reflected radio path is employed, that is, the reflected link capacity. Finally, the capacity of a backhaul link established between two IAB nodes $c$ and $d$ is denoted by the parameter $C_{(c,d)}^\text{BH}$.

We consider each TP as covered when a specific traffic demand can be guaranteed on both the direct and reflected link. In particular, the parameter $D$ represents the demand to be guaranteed by the line of sight connection with the BS. On the RIS-reflected link, a fraction $\xi \in [0,1]$ of the traffic demand $D$ is guaranteed. This represents the available backup link throughput, which can be activated when the primary link is experiencing outages, thanks to the continuous monitoring of the direct link during regular network operation. Whenever an obstruction is detected, the RIS of the SRC can be reconfigured to offer the alternative reflected path. In this case, the TP would experience a throughput degradation given by parameter $\xi$, but the connectivity would be preserved.
As previously mentioned, our model favours the activation of those SRCs offering high angular separation between the direct and reflected lines of sight. This is achieved by considering parameter $\Theta_{(t,c,d)}$, that denotes the smallest angle between CSs $c$ and $d$, as seen from TP $t$.
Parameter $L_{(t,c)}$, representing the distance between TP $t$ and CS $c$, is used to control the access link length.
Instantaneous reconfiguration of RIS reflecting elements is assumed, such that the surface can switch to assist different SRCs. However, we allow up to 1 TX-RX pair to be RIS-assisted at any time, leading to RIS being shared among SRCs on a time basis.
These previous assumptions may result to be technologically challenging for current RIS hardware manufacturers. However, we believe them to represent a realistic technological maturity level which should be considered for a robust investigation of potential RIS benefits.
RIS prototypes have presented a limited array field of view , similar to what happens for uniform planar antenna arrays~\cite{tan2018,beam_pattern}. Consequently, the orientation of installed RISs
needs to be such that the lines of sight of the base stations and test points of all SRCs to which the RIS is assigned fall inside the RIS field of view. In this work, we define a horizontal field of view angle $F$, and we discard the vertical\footnote{While a vertical field of view usually has a limited impact on the final network plan, it can be easily included in the model, if needed.} field of view.
Finally, the entire network plan is limited to a budget value $B$. Prices $P^\text{IAB},P^\text{RIS}$ represent the costs of installing an IAB and an RIS in any CS, respectively.

Here follow the decision variables on which the proposed planning model is based: 
\begin{itemize}
    \item $y_c^\text{DON}, y_c^\text{IAB}, y_c^\text{RIS}\in \{0,1\}$: installation variables equal to 1 if a donor, an IAB node or a RIS is installed in CS $c \in \C$, 0 otherwise,
    \item $x_{t,c,r} \in \{0,1\}$: SRC activation variable, equal to 1 if RIS in $r \in \C$ assists the communication between IAB node in $c \in \C$ and TP $t \in T$, 0 otherwise,
    \item $z_{c,d} \in \{0,1\}$: backhaul link activation equal to 1 if IAB node in $c \in \C$ is connected to IAB node $d \in \C$, 0 otherwise, 
    \item $f_{c,d} \in \mathbb{R}^+$: backhaul traffic flowing from IAB node $c \in \C$ to IAB node $d \in \C$,
    \item $t_{c}^\text{TX} \in [0,1]$: for any IAB node $c \in \C$, the fraction of time spent in transmission,
    \item $\delta_{r} \in [0,2\pi]$: RIS orientation variable, representing the surface azimuth with respect to a reference direction,
    \item $l_{t} \in \mathbb{R}^+$: average between direct (with IAB node) and secondary access (with RIS) link lengths covering TP $t \in \T$,
    \item $\theta_t \in [0,\pi]$ angular separation between direct and reflected links covering TP $t \in \T$.
\end{itemize}
With the given notation, parameters and decision variables, we now introduce a MILP (Mixed Linear Linear Programming) formulation of the IAB network planning problem with RISs:

{
\input{content/model_flexible}
}
The objective function~(\ref{opt:obj}) maximizes the average angular separation over the planned network while minimizing the average link length\cite{devoti2020}. In particular, parameters $\Theta$ and $\hat{L}$ are used to normalize the angular separation sum and link length sum over all TPs, while parameter $\mu \in [0,1]$ can be used to tune the maximization emphasis toward one of the two objectives. 
Constraint~(\ref{opt:ris_iab_act}) makes sure that IAB nodes and RISs are not installed in the same CS.
From a modelling standpoint, a donor is equivalent to an IAB node connected to the core network. Consequently, a donor can only be active in a candidate site populated by an IAB node, as stated by constraint~(\ref{opt:don_act}). Constraint~(\ref{opt:budget}) enforce a budget limit $B$ for the network deployment cost, which is given by the number of installed IAB nodes and RISs and the respective costs $P^\text{IAB}, P^\text{RIS}$.
Using binary link activation parameters $\Delta_{(t,c,r)}^\text{SRC}$, constraint~(\ref{opt:bh_link_act}) allows for an SRC variable to be active only if the necessary radio connections can be established and if the IAB node and RIS involved in the SRC are installed. Similarly, constraint~(\ref{opt:src_act}) allows for a backhaul link to be active only if the two nodes are installed, and a radio connection can be established between the two.
Constraint~(\ref{opt:one_src}) makes such that each TP is covered by one SRC.

According to standard specifications, IAB networks are expected to assume spanning-tree topology~\cite{3gpp.38.3874}. In our model, constraint~(\ref{opt:spanning_tree}) enforces the creation of a spanning tree by allowing for up to one ingress link per IAB node and none for the donor.
For each node, the traffic flow entering from the core network and the parent IAB node is balanced with the traffic leaving the node towards the associated UEs and downstream IAB nodes through constraint~(\ref{opt:flow_balance}). Note that the traffic coming from the core is available only at the donor, and it is equivalent to the single user demand $D$ times the number of test points $|\T|$. Additionally, the constraint above forces the activation of exactly one donor, as the traffic would be unbalanced otherwise. Constraint~(\ref{opt:flow_act}) is such that no traffic can flow on backhaul links which have not been activated; this also limits the flow to the backhaul link capacity $C_{(c,d)}^\text{BH}$.

In constraint~(\ref{opt:tx_time}) we set variable $t_c^\text{TX}$ to the fraction of time that each donor/IAB node spends during transmission. We assume that node resources are shared on a time-division basis (i.e., a node can transmit either to another node or to a UE at any moment) as specified by 3GPP Rel. 15. Consequently, quantity $t_c^\text{TX}$ is given by the sum of the time-fraction spent transmitting to downstream IAB nodes and the time-fraction spent transmitting to all the associated test points. The former is computed by considering the backhaul transmission airtime, namely the ratio between the flow $f_{c,d}$ and the link capacity $C_{(c,d)}^\text{BH}$. Similarly, the latter can be computed by considering the access transmission airtime. However, since the node can transmit to each test point through either the direct or reflected link, the longer between the direct and the reflected airtime is considered.
Constraint~(\ref{opt:tdm}) sums the transmission airtime variable set in the previous constraint with the time-fraction that the node spends in reception. The sum is then limited to 1, meaning that no more than 100\% of the node time-sharing resources can be occupied. Additionally, this enforces half-duplex for each node.
While each RIS can be part of multiple SRCs, we allow only one TX-RX pair to be served at any time by a single RIS. Each reconfigurable surface is thus shared among different SRCs through a time-division process, similarly to what is assumed for IAB nodes and donors. In the proposed planning formulation, this is achieved through constraint~(\ref{opt:ris_tdm_2}), where the cumulative airtime of TPs associated with a RIS is limited to 1.

Constraints from~(\ref{opti2:or1}) to~(\ref{opti2:or4}) set each RIS orientation such that the associated IAB nodes and UEs all fall inside its field of view $F$, which is achieved by means of parameters $\Phi_{(r,t)}^\text{A}$ and $\Phi_{(r,d)}^\text{B}$~\cite{Moro2021}. These represent the angles between the line connecting the RIS $r$ to TP $t$ or to CS $c$, respectively, and the reference direction.
Finally, constraint~(\ref{opt:ang_sep}) sets $\theta_t$ to the minimum angular separation between the line connecting the TP with the associated IAB node and the line connecting the TP with the associated RIS. Constraint~(\ref{opt:lin_len}) sets variable $l_t$ to the average between the IAB node-TP link length and the RIS-TP link length.

%% file: content/model_flexible.tex
{
\small
\begin{subequations}
\begin{equation}
\max \sum_{t \in \T} \left\{ \mu \sum_{t \in \T}\frac{\theta_t}{\Theta} - (1-\mu) \sum_{t \in \T} \frac{l_t}{\hat{L}} \right\}\label{opt:obj}\\
\end{equation}
\vspace{-0.7cm}
\begin{flalign}
&\text{s.t.:}&\nonumber\\
&y_c^\text{IAB} + y_c^\text{RIS}\leq 1 & \forall c \in \C,\label{opt:ris_iab_act}\\
&y_c^\text{DON} \leq y_c^\text{IAB}& \forall c \in \C,\label{opt:don_act}\\
%
%
&\sum_{c \in \C}\left( P^\text{IAB}y_c^\text{IAB} + P^\text{RIS}y_c^\text{RIS}\right) \leq B,&\label{opt:budget}\\
&z_{c,d} \leq \Delta^\text{BH}_{(c,d)}\left( y_c^\text{IAB} + y_d^\text{IAB} \right)/2,&\forall c,d \in \C,\label{opt:bh_link_act}\\
&x_{t,d,r} \leq \Delta^\text{SRC}_{(t,d,r)}\left( y_d^\text{IAB} + y_r^\text{RIS} \right)/2,& \forall t \in \T, d,r \in \C,\label{opt:src_act}\\
&\sum_{d,r \in C}x_{t,d,r} = 1,&\forall t \in T,\label{opt:one_src}\\
&\sum_{d \in \C}z_{d,c} \leq 1-y_c^\text{DON},&\forall c \in \C,\label{opt:spanning_tree}
\end{flalign}
\vspace{-0.5cm}
\begin{flalign}
&|\T|Dy_c^\text{DON} + \sum_{d \in \C}\left( f_{d,c} - f_{c,d} \right) - \sum_{\substack{t \in \T\\ r \in \C}}Dx_{t,d,r} = 0,&\forall c \in C,\label{opt:flow_balance}
\end{flalign}
\vspace{-0.7cm}
\begin{flalign}
%
%
&f_{c,d} \leq C_{(c,d)}^\text{BH}z_{c,d}, &\forall c,d \in \C,\label{opt:flow_act}
%
%
\end{flalign}
\vspace{-0.7cm}
\begin{flalign}
\footnotesize
&t_c^\text{TX} = \sum_{d \in \C} \frac{f_{c,d}}{C_{(c,d)}^\text{BH}} + \sum_{\substack{t \in \T\\r\in \C}} \max\left\{\frac{D}{C_{(t,d,r)}^\text{DIR}},\frac{\xi D}{C_{(t,d,r)}^\text{REF}}\right\}x_{t,d,r} &\hspace{-0.45cm}\forall c \in \C\hspace{-0.2cm}\label{opt:tx_time}
\end{flalign}
\vspace{-0.5cm}
\begin{flalign}
&\sum_{d \in \C} \frac{f_{d,c}}{C_{(d,c)}^\text{BH}} + t_c^\text{TX} \leq 1 &\forall c \in \C,\label{opt:tdm}\\
%
%
\end{flalign}
\begin{flalign}
&\sum_{\substack{t\in \T\\ d \in \C}}x_{t,d,r}\frac{\xi D}{C_{(t,d,r)}^\text{REF}} \leq 1,&\forall r \in\C,\label{opt:ris_tdm_2}\\
%
%
&\phi_r\geq \Phi^{\text{A}}_{(r,t)} - F/2 - 2\pi(1- x_{t,d,r})&\forall t \in \T, d,r \in \C,\label{opti2:or1}\\
&\phi_r\leq \Phi^{\text{A}}_{(r,t)} + F/2 + 2\pi(1- x_{t,d,r})&\forall t \in \T, d,r \in \C,\label{opti2:or2}\\
&\phi_r\geq \Phi^{\text{B}}_{(r,d)} - F/2 - 2\pi(1- x_{t,d,r})&\forall t \in \T, d,r \in \C,\label{opti2:or3}\\
 &\phi_r\leq \Phi^{\text{B}}_{(r,d)} + F/2 + 2\pi(1- x_{t,d,r})&\forall t \in \T, d,r \in \C,\label{opti2:or4}\\
&\theta_{t} \leq \Theta_{(t,c,r)} + 2 \pi(1-x_{t,c,r})&\forall t \in \T, c,r \in\C,\label{opt:ang_sep}\\
&l_t\geq \frac{1}{2}\sum_{c,r \in \C, v \in     V}x_{t,c,r} (L_{(t,c)} + L_{(t,r)})&\forall t \in \T.\label{opt:lin_len}
\end{flalign}
\end{subequations}
}

%% file: content/5_results.tex
In this section, we give a reliability analysis of HF RAN deployments that have been optimized employing the planning formulation shown in Section~\ref{sec:ris_model}. In particular, we compare the results given by RIS-enabled planning against a baseline approach where only donors and IAB nodes can be installed. This IAB-only formulation is available in the extended version of this work~\cite{fiore2021boosting}. In the baseline case, dual connectivity is still guaranteed by covering each TP with two different IAB nodes instead of relying on a RIS-reflected path. However, apart from the different radio link nature, all the planning assumptions, objectives and parameters remain unchanged.

The considered planning instances are characterized by a deployment area of $300\text{x}400m$ where 25 CSs can be activated, and 15 TPs need to be covered.
Donors and IAB nodes are modelled as 64-elements uniform linear array antennas with a transmission power of $30dBm$ and carrier frequency set to $28GHz$. Receivers are modelled as omnidirectional antennas. RISs are modeled as planar surfaces of $50\text{x}50cm$ containing $10^4$ passive reflecting elements, which is compatible with a $\lambda/2$ inter-element spacing~\cite{Renzo2020}.
The path-loss model of~\cite{Akdeniz2014} and the received power formulas in~\cite{Wang2019} were employed to precompute the average SNR of the links of every SRC combination. Finally, achievable rates were extracted according to NR modulation and coding scheme.
TP traffic demand over primary links is set to $100 Mbps$. The degradation factor $\xi$ is set to $0.5$, leading to a demand of $50Mbps$ guaranteed on each secondary link.
All the results in this section were obtained by averaging 20 random drops of CS and TP positions. Random instances were generated though MATLAB and planning optimization was solved by CPLEX.
\vspace{-0.1cm}
\subsection{Budget Sensitivity}
\begin{figure}[!t]
\centering
\subfloat[Average angular separation]{\includegraphics[width=1.6in]{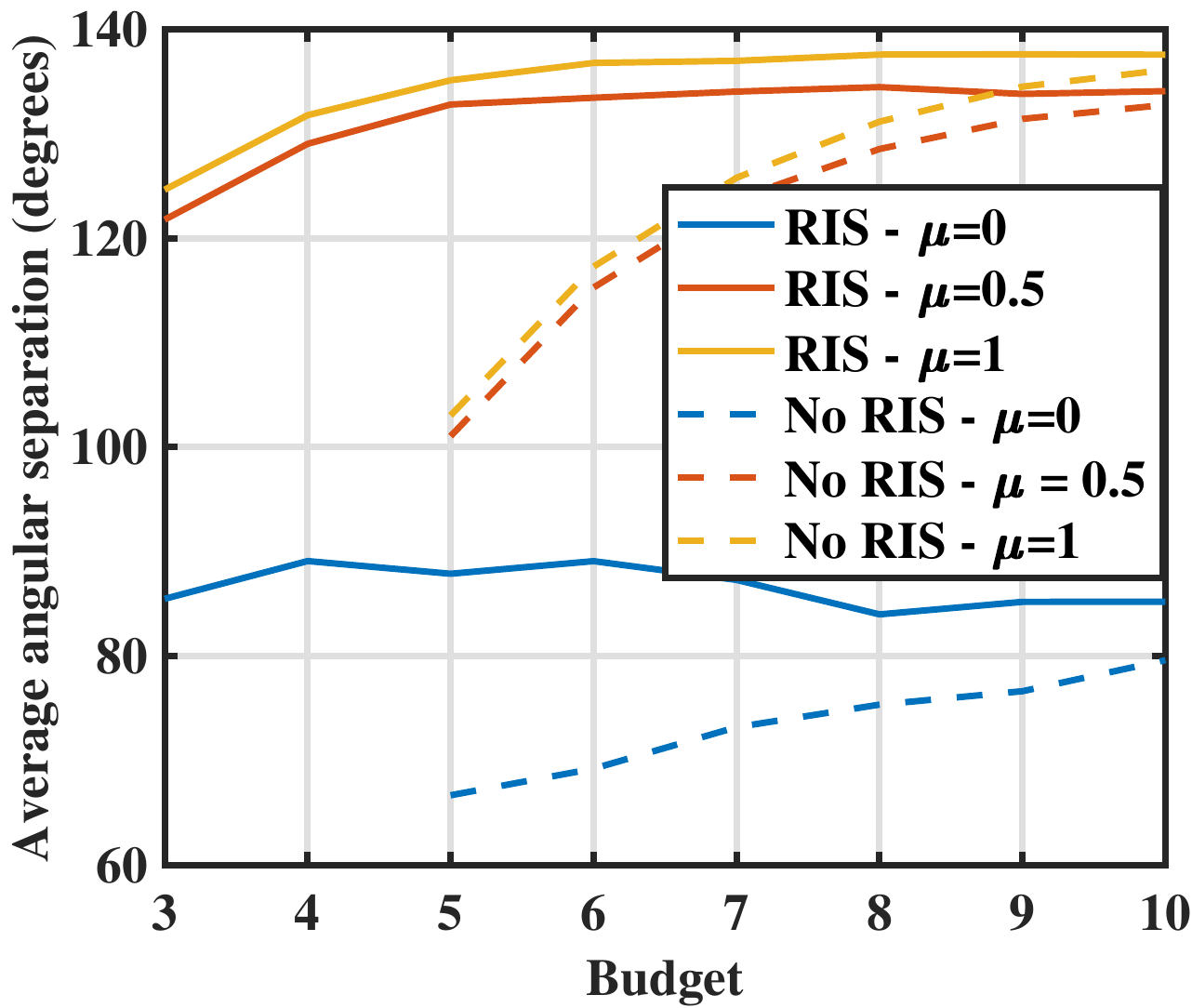}%
\label{fig:b_vs_angle}}
\hfil
\subfloat[Average SRC link length]{\includegraphics[width=1.6in]{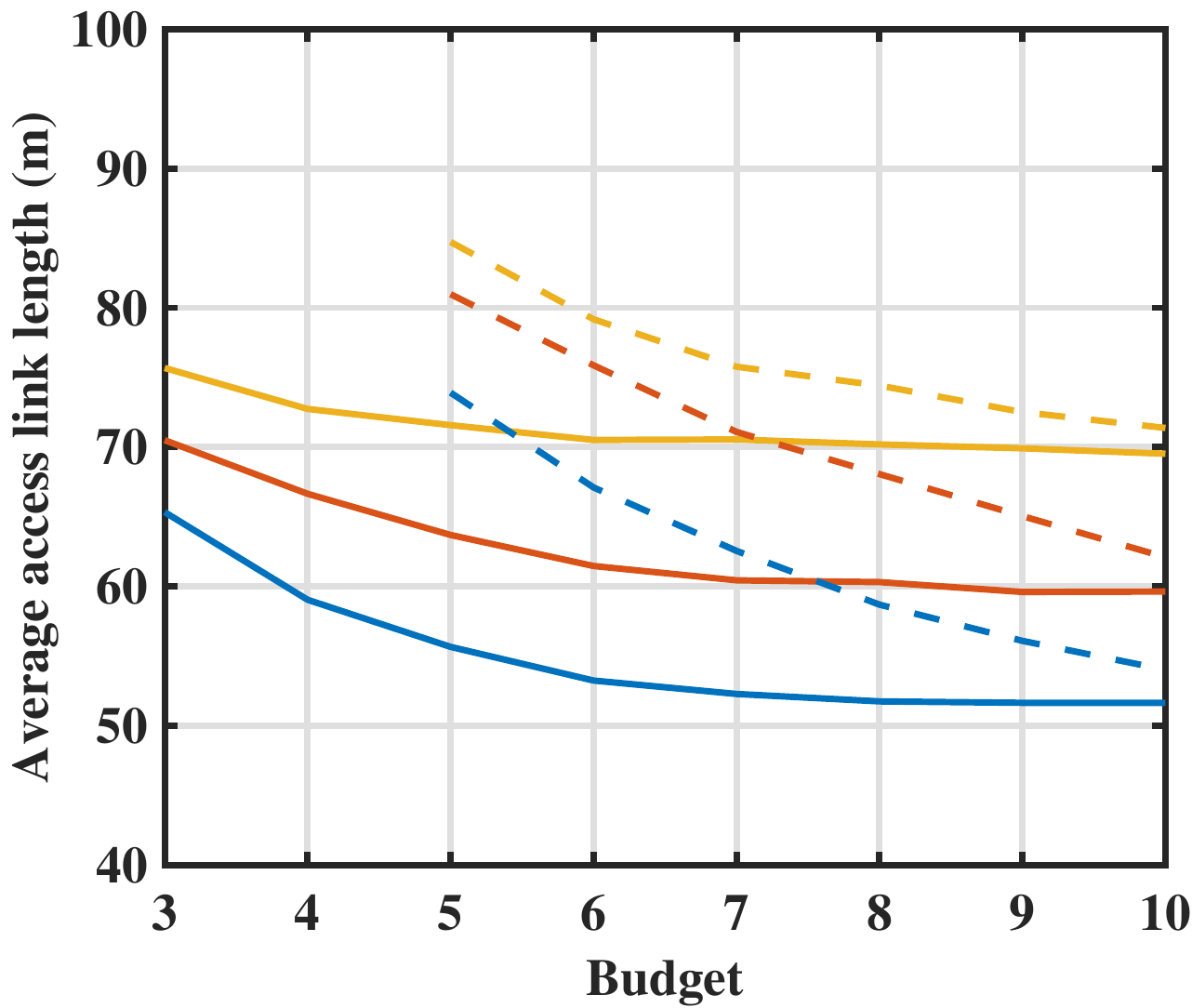}%
\label{fig:b_vs_l}}
\caption{Angular separation and link length sensitivity to budget variations.}
\label{fig:budget_vars}
\vspace{-5mm}
\end{figure}
As RISs are expected to have an impact on deployment costs, we carried out a budget sensitivity analysis by evaluating the blockage resilience indicators, i.e. average angular separation and link length, for different budget planning values. A clear indication of the cost of deploying a RIS has yet to be found in the literature. However, prototypes have shown how a RIS can be realized with inexpensive materials and production processes. Thus, an order of magnitude between the cost of IAB nodes and RIS seems reasonable. Consequently, we have set to 1 the cost of deploying an IAB node and to 0.1 the price of installing a RIS.
In Figures~\ref{fig:b_vs_angle} and~\ref{fig:b_vs_l} we plot the blockage resilience indicators against a planning budget that spans from 3 to 10 units.
In particular, in Figure~\ref{fig:b_vs_angle} we compare the average angular separation of RIS-enabled planning (solid lines) and IAB-only planning (dashed lines). The comparison is carried out by considering different values of $\mu$ in the objective function~(\ref{opt:obj}). Namely, we evaluated the results when either only the maximization of the average angular separation is considered ($\mu=1$), or when only the minimization of the link length is considered ($\mu=0$). Furthermore, we have analyzed $\mu=0.5$, which represents a balance between the two objectives. A similar analysis is shown in Fig.~\ref{fig:b_vs_l} for the average SRC link length. 
In both figures and for any considered $\mu$, the dotted lines, corresponding to the baseline approach, are interrupted for $B<5$. No solution could be found with these budget values. That is, a traditional HF RAN cannot guarantee the minimum rates when no RISs are installed. On the other hand, RIS-enabled planning is still feasible for these values, showing how RIS can be used to solve planning scenarios with strict budgets.
As the budget increases past 5 units, the baseline approach can be brought to a solution. However, RIS-enabled planning generally outperforms the IAB-only approach in terms of both angular separation and link length. Indeed, a budget increase of approximately 40\% is required such that the IAB-only model can reach the same level of angular separation and link length of RIS-enabled planning.
This behaviour is due to the reduced RIS installation costs that allow more RISs to be installed even with lower budgets. Indeed, in our results, the average number of installed IAB nodes for the baseline approach is always 1 to 3 nodes higher than the case when RIS are available. This suggests that, while the number of necessary IAB nodes can be reduced by employing RISs, these cannot be considered a complete substitute of IAB nodes since a considerable number of them need to be activated. Finally, for very large budget values, both RIS-enabled and traditional HF RAN approaches reach the same angular separation and link length level. This is an expected result since such budgets allow the activation of the optimal set of candidate sites that bring the aforementioned topological properties to their saturation level, independently on the technology employed for the backup link. In other words, when the deployment is not budget-restricted, there is no significant difference between installing RISs or IAB nodes.
\vspace{-0.1cm}
\subsection{Obstacle Deployment}
\begin{figure}[!t]
    \centering
    \subfloat[Percentage of served TPs]{\includegraphics[width=1.6in]{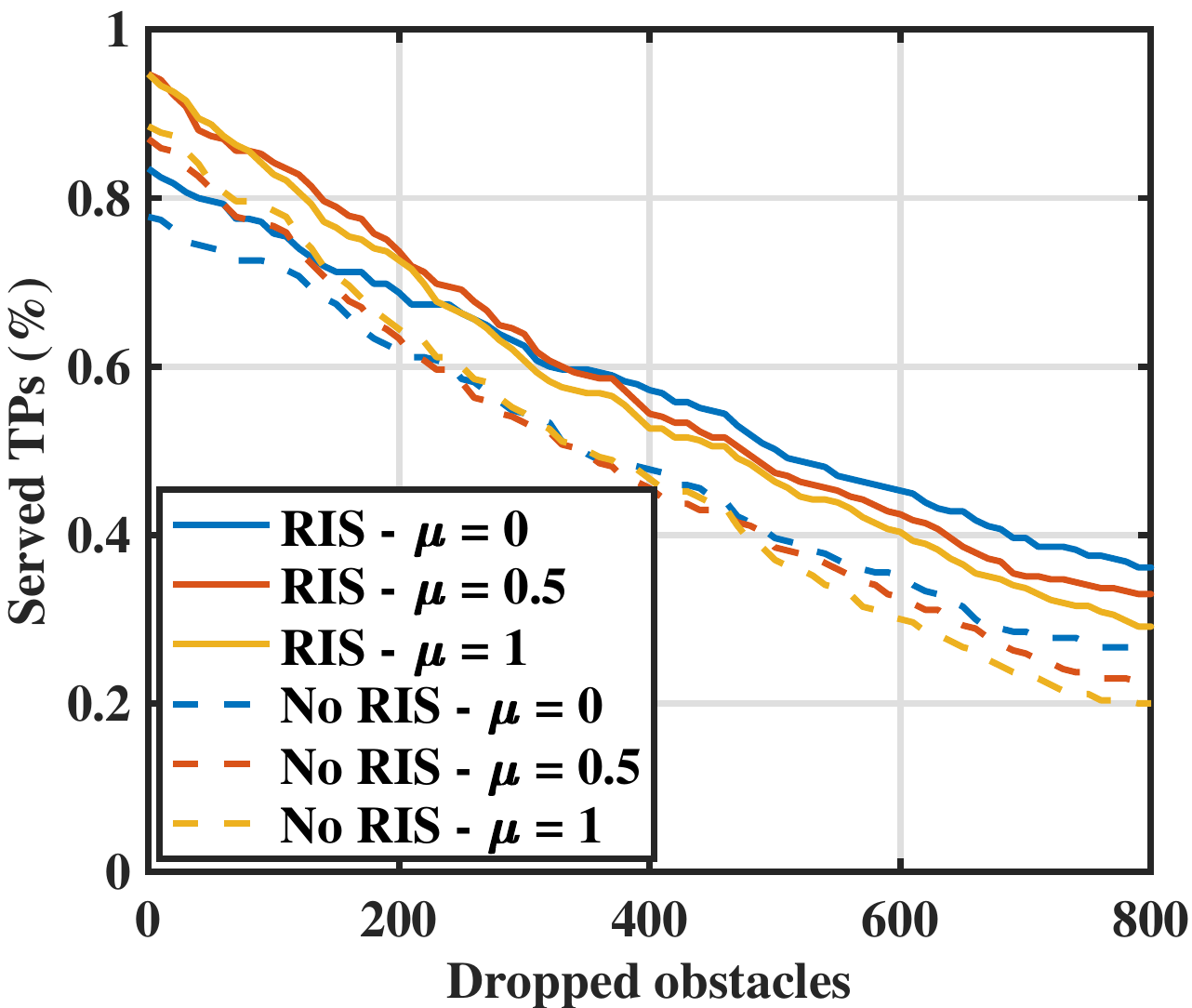}%
\label{fig:o_vs_tp_15}}
\hfil
\subfloat[Resilience gain]{\includegraphics[width=1.6in]{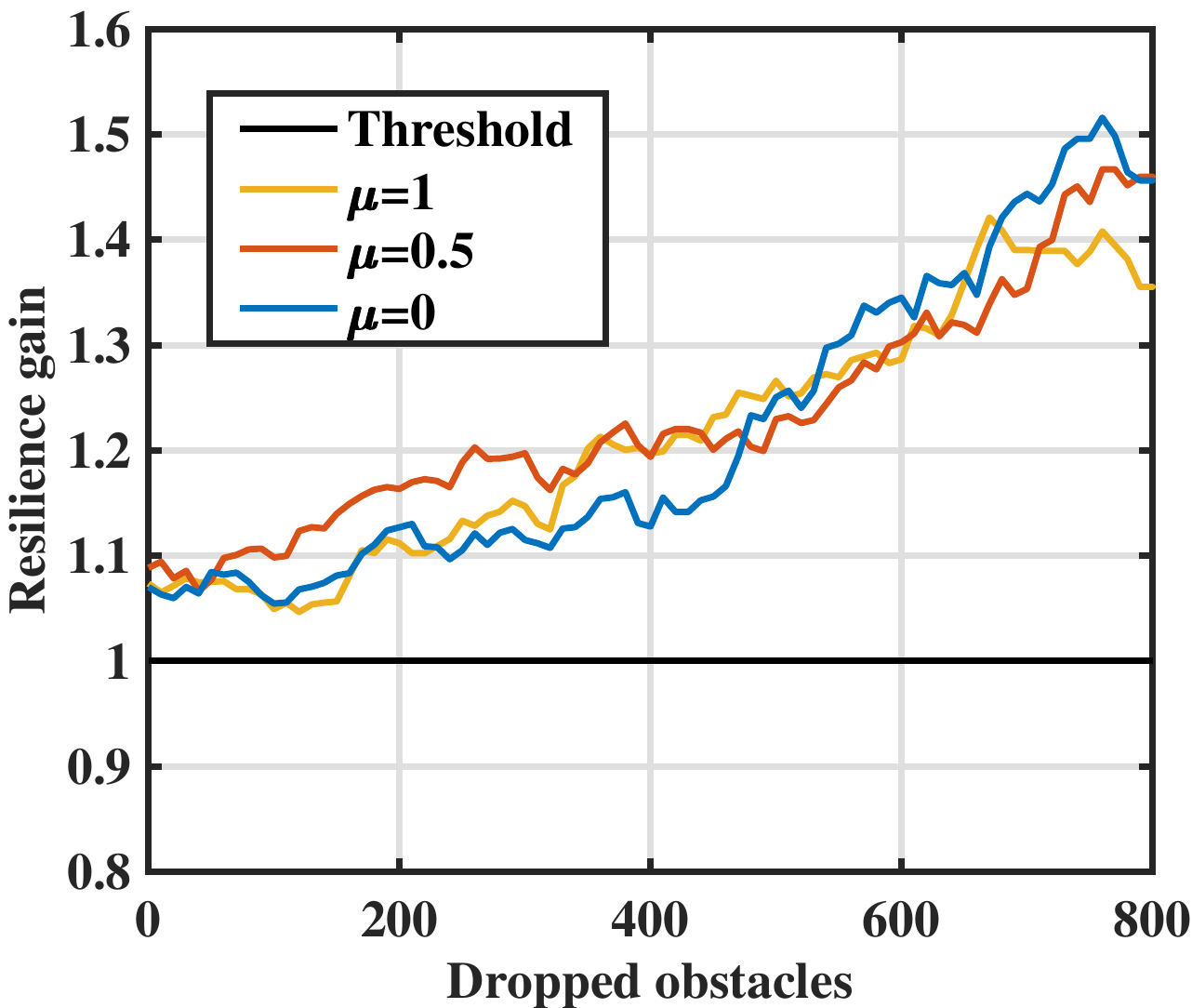}%
\label{fig:ratio_vs_tp_15}}
\caption{\footnotesize Impact of obstacles on TPs in outage, $B=5$.}
\label{fig:o15}
\vspace{-7mm}
\end{figure}

In the previous subsection, we have numerically evaluated the angular separation and link length of optimized RAN layouts. These topological properties, when properly optimized, are expected to give robustness to HF RAN. In this subsection, we enrich this analysis by evaluating how these properties impact the resilience of the resulting planned network during normal operating conditions. To do this, we drop randomly oriented, 5-meter-long linear obstacles into random positions of the planning area. Additionally, we have included a self-blockage angular sector, representing the human body obstruction to each TP. The angle span is either $2\pi/3$ or $8\pi/9$ with a probability of $0.5$ each, and it is randomly oriented~\cite{GRANW2018study}.
Any radio link that falls inside the self blockage sector or that is crossed by a randomly dropped obstacle is considered blocked.
Figure~\ref{fig:o_vs_tp_15} shows the percentage of served TPs (i.e., at least one between primary and secondary link is not blocked) for an increasing number of dropped obstacles and a planning budget of 5 units. In general, the baseline approach (dashed lines) is consistently outperformed by the RIS-enabled (solid lines), confirming the results of the previous paragraph. Indeed, Figure~\ref{fig:ratio_vs_tp_15} shows that RIS-enabled planning results in up to 50\% more served TPs with respect to the baseline approach. 
Additionally, Figure~\ref{fig:o_vs_tp_15} shows how the link length-minimizing objective function ($\mu=0$) outperforms the other two objective approaches 
as soon as the dropped obstacles increase past 300. This suggests that the planning objective should be chosen according to the expected obstacle density.
\begin{figure}[!t]
    \centering
    \subfloat[Percentage of served TPs]{\includegraphics[width=1.6in]{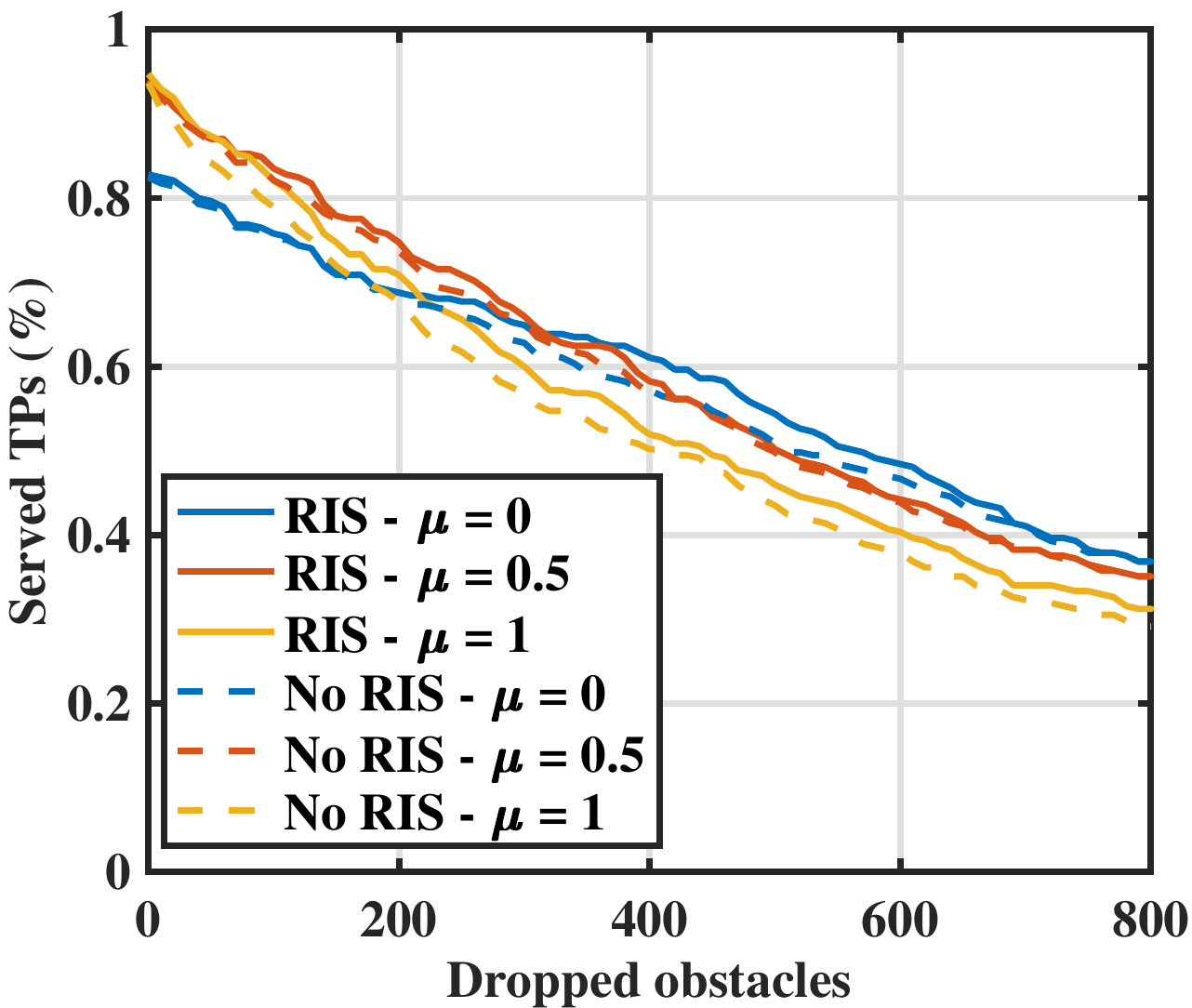}%
\label{fig:o_vs_tp_20}}
\hfil
\subfloat[Resilience gain]{\includegraphics[width=1.6in]{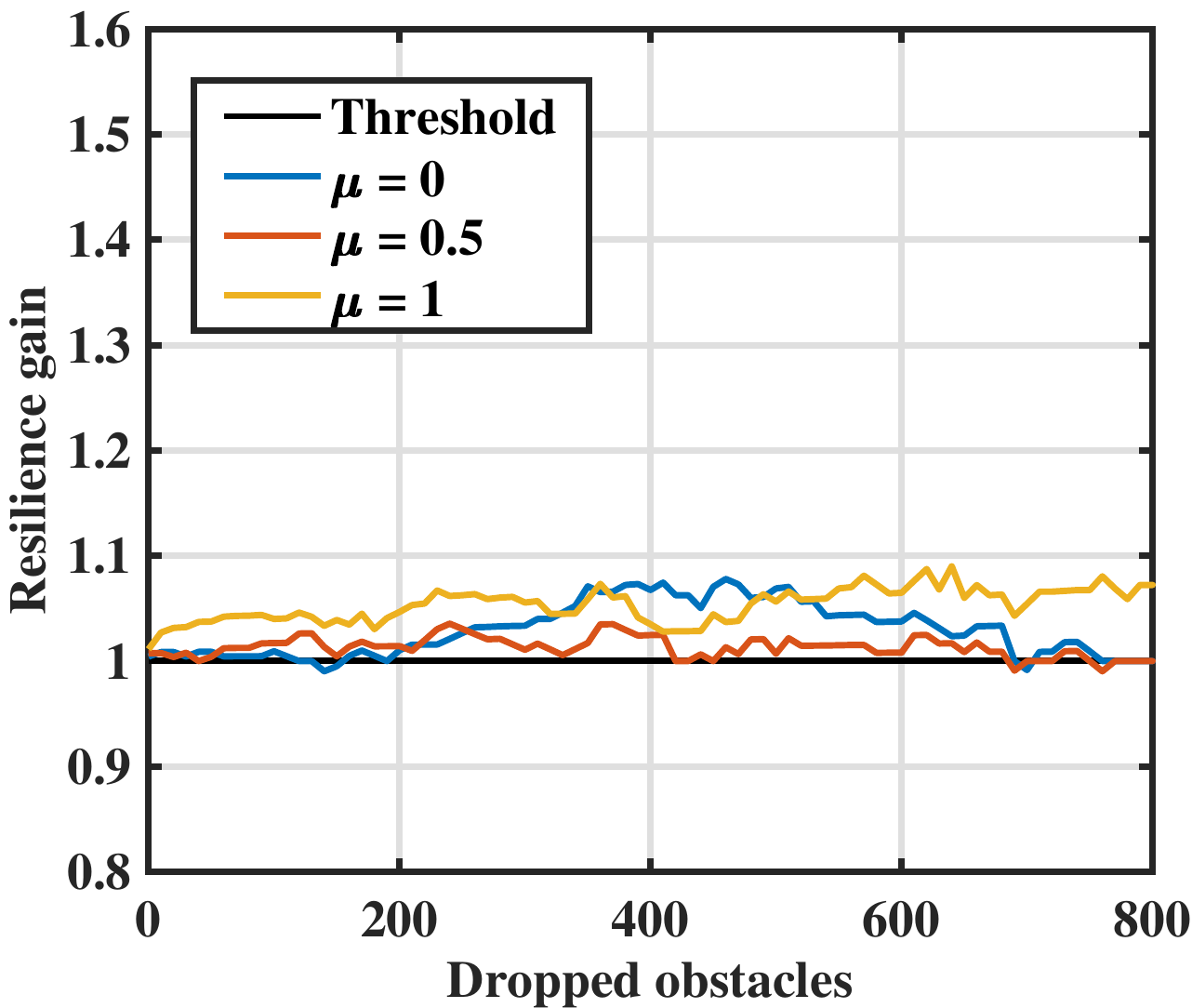}%
\label{fig:ratio_vs_tp_20}}
\caption{\footnotesize Impact of obstacles on TPs in outage, $B=10$.}
\label{fig:o20}
\vspace{-6mm}
\end{figure}
As previously mentioned, higher budget levels allow for the installation of more IAB nodes, and the objective function values of the baseline converge to those of the RIS-enabled planning. Consequently, the resilience gain of the RIS-based approach is shortened for $B=10$, as shown in Figure~\ref{fig:o20}.
In particular, Figure~\ref{fig:o_vs_tp_20} shows a reduced difference between the percentage of served TPs of the baseline and RIS-enabled approach when the budget value is set to $10$ units. Similarly, the resilience gain of using RISs with $B=10$, in Figure~\ref{fig:ratio_vs_tp_20}, is limited to a maximum of 10\%.
Finally, these results confirm again how RISs are better employed in budget-restricted planning scenarios, as they effectively provide cost-effective blockage resilience through link densification.

%% file: content/6_conclusion.tex
We have analyzed the resilience of optimized mm-Wave IAB RAN deployments with RISs. To do so, we have developed a novel MILP network planning model that privileges topological characteristics such as angular separation and link length, proven to be effective in enhancing RAN reliability against random obstacles. In our evaluation campaigns, we have compared the results of the RIS-enabled planning against a baseline approach where no RIS is installed. Numerical results showed an increase in access resilience and a reduction in budget expenditure through the usage of RISs, compared to the more traditional IAB-only RAN.

%% file: content/7_additional_material.tex
We hereby detail the planning formulation for an HF RAN where each test point is associated with a primary and secondary IAB node. This formulation was employed to produce baseline results in Section~\ref{sec:results} and it is based on the same assumptions and parameters described in Section~\ref{sec:ris_model}.\newline 
In this formulation, parameter $U^\text{WRD}$ represents the capacity of the wired connection between donors and the core network, which is assumed to be infinite. Parameter $\Delta_{(t,d)}^\text{ACC}$ expresses the possibility of activating an access link between TP $t\in \T$ and CS $c \in \C$.\newline
The optimization variables are the following: 
\begin{itemize}
    \item $y_c^\text{DON}, y_c^\text{IAB} \in \{0,1\}$: installation variables equal to 1 if a donor or an IAB node is installed in CS $c \in \C$, 0 otherwise,
    \item $x_{t,c} \in \{0,1\}$: primary access link activation variable, equal to 1 if test point $t \in \T$ is covered by a primary link with donor/IAB node in CS $c\in \C$, 0 otherwise,
    \item $s_{t,c} \in \{0,1\}$: backup access link activation variable, equal to 1 if test point $t \in \T$ is covered by a backup link with donor/IAB node in CS $c\in \C$, 0 otherwise,
    \item $z_{c,d} \in \{0,1\}$: backhaul link activation equal to 1 if BS in $c \in \C$ is connected to BS $d \in \C$, 0 otherwise, 
    \item $f_{c,d} \in \mathbb{R}^+$: backhaul traffic flowing from BS $c \in \C$ to BS $d \in \C$,
    \item $w_{c} \in \mathbb{R}^+$: traffic flowing from the core network into donor $c \in \C$ through a wired connection,
    \item $t_{c}^\text{TX} \in [0,1]$: for any BS $c \in \C$, the fraction of time spent in transmission,
    \item $l_{t} \in \mathbb{R}^+$: average between primary and backup access link lengths covering TP $t \in \T$,
    \item $\theta_t \in [0,\pi]$ angular separation between primary and backup links covering TP $t \in \T$.
\end{itemize}
Here follows the MILP formulation used to optimize HF IAB RAN deployments:
\begin{subequations}
\begin{equation}
\max \sum_{t \in \T} \left\{ \mu \sum_{t \in \T}\frac{\theta_t}{\Theta} + (1-\mu) \sum_{t \in \T} \frac{l_t}{\hat{L}} \right\}\label{opt2:obj}
\end{equation}
\vspace{-0.5cm}
\begin{flalign}
&\text{s.t.:}&\nonumber\\
&y_c^\text{DON} \leq y_c^\text{IAB}& \forall c \in \C,\label{opt2:don_act}\\
 &\sum_{c \in \C}y_c^\text{DON}\leq 1,&\label{opt2:only_1_don}\\
&\sum_{c \in \C} P^\text{IAB}y_c^\text{IAB} \leq B,&\label{opt2:budget}\\
&z_{c,d} \leq \Delta^\text{BH}_{c,d}\left( y_c^\text{IAB} + y_d^\text{IAB} \right)/2,&\forall c,d \in \C,\label{opt2:bh_link_act}\\
&x_{t,d} \leq \Delta^\text{ACC}_{(t,d)}y_d^\text{IAB},& \forall t \in \T, d \in \C,\label{opt2:main_act}\\
&s_{t,d} \leq \Delta^\text{ACC}_{(t,d)}y_d^\text{IAB},& \forall t \in \T, d \in \C,\label{opt2:backup_act}\\
&\sum_{d \in C}x_{t,d} = 1,&\forall t \in T,\label{opt2:one_main}\\
&\sum_{d \in C}s_{t,d} = 1,&\forall t \in T,\label{opt2:one_backup }\\
&\sum_{d \in \C}z_{d,c} \leq 1-y_c^\text{DON},&\forall c \in \C,\label{opt2:spanning_tree}
\end{flalign}
\vspace{-0.5cm}
\begin{flalign}
w_c + \sum_{d \in \C}\left( f_{d,c} - f_{c,d} \right) - \sum_{t \in \T}(Dx_{t,c}+\xi D s_{t,c})= 0,&\forall c \in C,\label{opt2:flow_balance}
\end{flalign}
\vspace{-0.5cm}
\begin{flalign}
&w_c \leq U^\text{WRD}y_c^\text{DON},&\forall c \in \C,\label{opt2:wired_cap}\\
&f_{c,d} \leq C_{(c,d)}^\text{BH}z_{c,d}, &\forall c,d \in \C,\label{opt2:flow_act}
\end{flalign}
\begin{flalign}
&t_c^\text{TX}=\sum_{d \in \C}\frac{f_{c,d}}{C_{(c,d)}^\text{BH}}  + \sum_{t \in \T} \left(\frac{D}{C_{(t,c)}^\text{ACC}}x_{t,c} + \frac{\xi D}{C_{(t,c)}^\text{ACC}}s_{t,c}\right),&\forall c \in \C,\label{opt2:tx_time}
\end{flalign}
\vspace{-0.5cm}
\begin{flalign}
&t_c^\text{TX} + \sum_{d \in \C}\frac{f_{d,c}}{C_{(d,c)}^\text{BH}} \leq 1,&\forall c \in \C, \label{opt2:tdm}\\
%
%
&\theta_{t} \leq \Theta_{(t,c,r)} + 2 \pi(2-x_{t,c} - s_{t,r})&\forall t \in \T, c,r \in\C,\label{opt2:ang_sep}\\
&l_t\geq \frac{1}{2}\sum_{c,r \in \C}\left(x_{t,c} L_{(t,c)} + s_{t,r}L_{(t,r)}\right)&\forall t \in \T.\label{opt2:lin_len}
\end{flalign}
\end{subequations}